# Chemical Potential Fluctuations in Topological Insulator $(Bi_{0.5}Sb_{0.5})_2Te_3$ Films Visualized by Photocurrent Spectroscopy

Christoph Kastl,[1,2] Paul Seifert,[1,2] Xiaoyue He,[3] Kehui Wu,[3] Yongqing Li,[3] Alexander Holleitner[1,2]*

[1] Walter Schottky Institut and Physik-Department, Technische Universität München, Am Coulombwall 4a, 85748 Garching, Germany.

[2] Nanosystems Initiative Munich (NIM), Schellingstr. 4, 80799 München, Germany.

[3] Institute of Physics, Chinese Academy of Sciences, Beijing 100190, China.

**Abstract**

We investigate the photocurrent properties of the topological insulator $(Bi_{0.5}Sb_{0.5})_2Te_3$ on $SrTiO_3$-substrates. We find reproducible, submicron photocurrent patterns generated by long-range chemical potential fluctuations, occurring predominantly at the topological insulator/substrate interface. We fabricate nano-plowed constrictions which comprise single potential fluctuations. Hereby, we can quantify the magnitude of the disorder potential to be in the meV range. The results further suggest a dominating photo-thermoelectric current generated in the surface states in such nanoscale constrictions.

In recent years, a class of solid state materials, called three-dimensional topological insulators, has emerged [1-3]. In the bulk, a topological insulator behaves like an ordinary insulator with a band gap. At the surfaces, topologically non-trivial, gapless states exist showing remarkable properties such as a helical Dirac dispersion and the suppression of backscattering of spin-polarized charge carriers [4, 5]. In principle, thin films of topological insulators have two surfaces; the bottom one facing the underneath substrate and the top one. Predominantly, the top surface states have been experimentally characterized. For instance, due to a penetration depth of ~3 nm, an angle resolved photoemission spectroscopy (ARPES) is mainly limited to the topological insulator - vacuum interface [4, 6-8]. This top surface is also approached by high-resolution scanning tunneling microscopes [5, 9-11], and it is always exposed to the environmental conditions of the experiments. However, the prototypical topological insulator compounds bismuth telluride, bismuth selenide and their respective alloys are not chemically inert under ambient conditions [12-14]. This can have an adverse impact on the topological states, because an interfacial inversion layer can form with trivial electronic properties [15]. In contrast, the bottom surface states are buried, encapsulated and therefore protected against degradation, which is an essential prerequisite for future electronic devices based on topological insulators.

We demonstrate that a photocurrent spectroscopy allows addressing electronic states at the topological insulator/substrate interface, comprising both surface and bulk states. We verify this scheme in circuits of thin $(Bi_{0.5}Sb_{0.5})_2Te_3$-films on $SrTiO_3$-substrates. Such substrates with a large dielectric constant are recently discussed to allow a reduction of potential fluctuations within two-dimensional materials [16]. We apply a so-called dynamic plowing lithography based on an atomic force microscope (AFM) to the $(Bi_{0.5}Sb_{0.5})_2Te_3$-films. Therewith, we thin down the films by a few nanometers and in turn, record a two-dimensional photocurrent map. The maps exhibit reproducible photocurrent patterns with positive and negative amplitudes and with a lateral extension of up to micrometers. We observe that the patterns are not changed, when the film-height is reduced by only a few nanometers, and that they are even stable after several cooling cycles, exposures to atmospheric conditions, and/or lithography steps. We interpret these findings in a way that the photocurrent patterns arise from long-range potential fluctuations at the interface between the topological insulator and the substrate. The observed photocurrent amplitudes are maximum when the Fermi-level is tuned to the charge neutrality point, which is consistent with a reduced screening of the potential fluctuations at a low electron density. Additionally, we laterally pattern the $(Bi_{0.5}Sb_{0.5})_2Te_3$-films into narrow constrictions by the use of the nano-plowing lithography. Such circuits effectively comprise single potential fluctuations. Corresponding photocurrent maps allow us to quantify the magnitude of the effective disorder potential to be in the order of meV. The results suggest that a photo-thermoelectric current within the surface states dominates the photocurrent in such narrow constrictions. We further discuss a photovoltaic current within the bulk states and the impact of lateral boundaries on the electrostatic potential landscape. Our photocurrent experiments reveal the possibility to selectively probe the transport properties of electronic states buried at the interface between a topological insulator and an insulating substrate, and to address surface states in narrow constrictions made out of topological insulators. Hereby, our results may prove essential for the integration of topological insulators in nanoscale optoelectronic circuits. The experiments further disclose the impact of potential fluctuations on the transport properties of topological insulators. This finding may have implications for optimizing the growth of thin topological films as well as for the interpretation of linear magneto-resistance data in terms of electrostatic disorder [17, 18].

Thin films of $(Bi_{0.5}Sb_{0.5})_2Te_3$ with a thickness of 15 nm (~15 quintuple layers) are grown by molecular beam epitaxy on a $SrTiO_3$(111)-substrate (Supplementary data) [19, 20]. Ternary topological insulators such as $(Bi_{1-x}Sb_x)_2Te_3$ or $Bi_2Te_2Se$ show substantially reduced bulk conductivity [6, 21-22] compared to simple binary compounds as for example $Bi_2Se_3$, $Sb_2Te_3$, and $Bi_2Te_3$. The latter exhibit an inherent bulk doping [17, 23]. The combination of $(Bi_{0.5}Sb_{0.5})_2Te_3$ as a ternary, bulk insulating topological insulator and $SrTiO_3$ as an ultrahigh-$k$ dielectric substrate allows for a full range control of the chemical potential from $p$-doping to $n$-doping via electrical back-gating at low temperatures [21]. The films are patterned into 50 μm wide Hall bar devices using standard optical lithography and plasma etching. Ohmic contacts to the films as well as a gate electrode at the backside of the substrates are formed by metallization of chromium and gold. Photocurrent measurements are carried out at temperatures 4.2 K $< T <$ 77 K using a confocal scanning laser microscope with an excitation wavelength $\lambda = 806$ nm and a diffraction limited spatial resolution of ~900 nm. The current is measured between electrically unbiased contacts, named source and drain, using a current-voltage converter (Figure 1a). By scanning the laser spot across the sample, the photocurrent $I_{\text{photo}}(x, y)$ is detected for each position on the sample. Figure 1b shows such a two-dimensional photocurrent map for the region marked by the dashed rectangle in Figure 1a. We find reproducible submicron photocurrent patterns with positive (red) and negative (blue) amplitudes. A positive (negative) amplitude corresponds to a current flowing from source to drain (drain to source). The photocurrent patterns stand in contrast to photocurrents routinely observed in planar circuits based on low-dimensional materials such as graphene, $MoS_2$ or semiconducting nanowires. In such nano-circuits, the photoresponse is typically dominated by photovoltaic and photo-thermoelectric currents generated at interfaces which are for example metal interfaces at ohmic contacts [24, 25], step edges of ultrathin layered materials [26, 27] as well as heterojunctions and built-in fields induced by electrostatic or chemical doping [28-30]. Here, rather than being localized near metal contacts, the photoresponse comprises complex spatial patterns spanning across the entire Hall bar.

We excite the $(Bi_{0.5}Sb_{0.5})_2Te_3$ with a photon energy of 1.54 eV ($\lambda = 806$ nm), which creates electron-hole pairs above the band-gap of $E_{\text{gap}} \sim$ 0.2 eV. Time-resolved angle resolved photoemission spectroscopy on topological insulators shows that high energy photogenerated charge carriers relax within fs to a heated Fermi-Dirac distribution via carrier-carrier and optical phonon scattering.

Afterwards the non-equilibrium Fermi-Dirac distribution equilibrates with the lattice on a ps-time scale [31, 32]. Then, the local photocurrent $\boldsymbol{j}_{\text{photo}}(x,y)$ can arise either due to a photovoltaic effect or due to a photothermoelectric effect. In the former case, photogenerated electrons and holes are spatially separated resulting in a current density $\boldsymbol{j}_{PV}$. In turn, the local current density generates an electric field $\boldsymbol{E}_{PV} = -\rho \boldsymbol{j}_{PV}$ where $\rho$ is the local resistivity [30]. In the latter case, the laser induced heat gradient $\nabla T$ produces a thermoelectric field $\boldsymbol{E}_{PTE} = -S\nabla T$ with the Seebeck coefficient $S$ [26]. As will be discussed in this letter, the photocurrent profile provides information on the local potential landscape of the $(Bi_{0.5}Sb_{0.5})_2Te_3$-films. Circular and linear photogalvanic effects are negligible for the present experiments, because the photo-excitation occurs at a normal angle of incidence [32]. We point out that the photocurrent amplitudes depend on the device geometry. For the region denoted as "I" and "II" in Figure 1b, the mean amplitude, i.e. the strength of the photoresponse, is approximately constant. However, as the channel widens, the current amplitude decreases. This becomes evident at the transition between the regions "II" and "I". The geometry-dependent reduction of the amplitude is discussed in detail in ref.[33] for $Bi_2Se_3$-based circuits. It can be understood as a decreased density of current stream lines that connect the local photocurrent to the contacts [34]. In the following, we elaborate that chemical potential fluctuations at the interface between the $(Bi_{0.5}Sb_{0.5})_2Te_3$-films and the $SrTiO_3$-substrates give rise to the observed photocurrent patterns.

We use the tip of an atomic force microscope (AFM) and employ a so-called dynamic plowing lithography to thin down the $(Bi_{0.5}Sb_{0.5})_2Te_3$-films with nm-precision (Figure 2a and Supplementary Note 2) [35, 36]. In tapping mode, the external drive amplitude of the AFM-tip is increased and the set-point of the feedback loop is decreased such that the interaction between the tip and the films is sufficient to mechanically remove sample material from the top surface of the $(Bi_{0.5}Sb_{0.5})_2Te_3$. Although initially proposed for structuring soft materials such as polymers, the AFM-lithography can equally be applied to semiconductors, metals, and topological insulators by using specially coated tips [35, 37]. The $(Bi_{0.5}Sb_{0.5})_2Te_3$-films have a pristine thickness of $z$ = 15 nm (~15 quintuple layers). They are successively thinned down by $\Delta z$ = 5 nm and $\Delta z$ = 9 nm for a lateral extension of ~10 μm (lower panels of Figures 2b, 2c, and 2d). The main graphs of Figure 2c and 2d show the corresponding photocurrent maps taken after the lithography steps. Intriguingly, the photocurrent patterns of the film thinned by $\Delta z$ = 5 nm (Figure 2c) are identical to the ones of the pristine film (Figure 2b). Consequently, the data suggest that the chemical potential fluctuations are independent of the top surface morphology and that the photocurrent mainly stems from the electronic states located at the bottom interface between the substrate and the topological insulator film. We further note that in-between the photocurrent measurements shown in Figure 2, the films are exposed to ambient conditions and temperature cycles. Hereby, the sequence of experiments depicted in Figure 2b and 2c demonstrates that the optoelectronic response near the bottom interface is independent of the environmental conditions.

When the film thickness is reduced to 6 nm (in between the triangles in Figure 2d), the photocurrent response changes. At the vertical step edges (blue and red triangle in Figure 2d), the photoresponse is enhanced by approximately one order of magnitude. Additionally, the sign of the current is reversed at the opposite step edges. The findings are consistent with a recent study, where a photo-thermoelectric effect was demonstrated at step edges of 4 - 12 quintuple layers (4 nm - 12 nm) thin $Bi_2Te_3$- and $Sb_2Te_3$-platelets [38]. In this interpretation, the photo-thermoelectric current results from the thickness dependence of the electronic band structure due to confinement effects. In addition, defect states and imperfections in the nano-plowed top surface may further influence the band structure. We expect a substantial quantum mechanical coupling due to hybridization of the top and bottom surface states only when the thickness of the $(Bi_{0.5}Sb_{0.5})_2Te_3$-films is less than 3-4 nm. We note, however, that for thicknesses below 6 nm, we find that the dynamic plowing lithography rather destroys the films, i.e. the films rupture into parts (data not shown).

The data shown in Figure 2 are measured at zero gate voltage. Figures 3a-c depict photocurrent maps as the Fermi-level is tuned from the *p*-type into the *n*-type regime via the backgate voltage. The laser excitation occurs in the center of the pristine Hall bar where the geometry of the sample does not constrain the photocurrent. The exact form of the photocurrent patterns depends on the applied backgate voltage, which suggests that they are caused by electrostatic potential fluctuations within the $(Bi_{0.5}Sb_{0.5})_2Te_3$-films. At $V_{max} \approx 10$ V (Figure 3b), the mean amplitude of the photoresponse is maximum. This is even better visible in the top graph of Figure 3d, which depicts the mean photocurrent amplitude $\langle |I_{photo}| \rangle$ as a function of the gate voltage. Each data point is calculated by averaging the absolute value of the current $|I_{photo}|$ for an entire two-dimensional photocurrent map. In particular, the photoresponse is enhanced near the charge neutrality point, where the overall conductance exhibits a minimum ($V_{min}$ in Figure 3d). In our interpretation, the enhancement of the photocurrent near the charge neutrality point can have two origins. On the one hand, the maximum of $\langle |I_{photo}| \rangle$ in Figure 3d can be related to the poor screening near the charge neutrality point of the bottom surface states, since a reduction of the screening leads to larger fluctuations in the effective electrostatic potential [39-41]. On the other hand, both the Seebeck coefficient $S$ as well as the local resistivity $\rho$ show an enhancement near the charge neutrality point which can effectively lead to an overall increased photoresponse of the two-dimensional $(Bi_{0.5}Sb_{0.5})_2Te_3$-film [30, 42].

In a next step, we focus on single potential fluctuations to uncover the dominating photocurrent mechanism within the bottom surface states. In particular, we create a constriction parallel to the sample boundary by completely removing the $(Bi_{0.5}Sb_{0.5})_2Te_3$-film along a thin straight line (Figure 4a and Supplementary data). In this way, a circuit is formed consisting of pristine two-dimensional source and drain regions as well as the narrow constriction. With a lateral width $d = (1.8 \pm 0.3)$ μm, the constriction comprises a sequence of single potential fluctuations (Figure 4b). In Figure 4c, the photocurrent amplitude vs. the gate voltage is plotted for each excitation position along the constriction. In contrast to the data presented in Figure 3, all photocurrent puddles change sign at a gate voltage $V_{zero}$. Notably, $V_{zero}$ is not constant along the constriction (Figure 4d). Therefore, we interpret $V_{zero}(x)$ to mimic the variation of the local chemical potential [24], since the Fermi-level can be tuned by the gate voltage as $E_F \sim (V_{gate} - V_{Dirac})^{1/2}$, with $V_{Dirac}$ the gate voltage at which the Fermi-level aligns with the Dirac point of the bottom surface state [39]. Within a plate capacitor model for the gate capacitance, the maximum peak-to-peak voltage $\Delta V_{zero}$ of ~9 V in Figure 4d translates to a value on the order of 6 meV for the potential fluctuations in the $(Bi_{0.5}Sb_{0.5})_2Te_3$-film.

The macroscopic Hall bar geometry does not provide the possibility to directly measure the constriction's conductance by a four-terminal configuration. Therefore, we combine different two-terminal configurations of the Hall bar to indirectly extract the individual dark conductance of the source contact and the one of the AFM-fabricated constriction as a function of $V_{gate}$ (see Figure 4e). We observe that the minimum conductance of the constriction is shifted to a more negative gate voltage by $\Delta V_{gate} \sim 90\ V$ compared to the pristine source contact. The shift suggests that at the constriction, the energetic position of the electronic states is altered. As a consequence, also the Dirac point of the surface state will be shifted downwards in energy. It is well known that plasma etching induces defects at unprotected surfaces [43, 44]. For the $(Bi_{0.5}Sb_{0.5})_2Te_3$-films, such defects could occur at the vertical facets of the lateral edges. As was recently proposed [45], the defects may result in a band bending. This explains the observed voltage shift and therefore energy shift of the electronic states at the edges of the $(Bi_{0.5}Sb_{0.5})_2Te_3$-films. We point out that we observe a similar shift when the constrictions are fabricated at the center of the Hall-bar by the dynamic plowing lithography (Supplementary data). For such constrictions, all edges are fabricated by the AFM lithography. Hereby, we conclude that the occurrence of defect states and their influence on the optoelectronic properties of a topological insulator are generic to nanofabricated circuits.

We model the optoelectronic response of the bottom surface states by a photothermoelectric current. Starting with the extracted dark conductance $G$ of the constriction (Figure 4e), we calculate the gate-voltage dependence of the Seebeck coefficient according to the Mott formula [42, 46]

$$S = -\frac{\pi^3}{3}\left(\frac{k_B^2 T}{e}\right) \cdot \frac{1}{G} \cdot \frac{dG}{dE_F}, \tag{1}$$

assuming a Dirac system with $v_{Dirac} = 3.5 \cdot 10^5 \mathrm{ms}^{-1}$ [47]. The result is presented in Figure 4f. In our interpretation, there exists one dominating potential fluctuation for each excitation position. Therefore, $S(V_{gate})$ is shifted by $\Delta V_{zero}$ when the laser is scanned along the constriction. Then, a laser induced heat gradient generates a thermovoltage due to the difference in thermopower along the constriction. The thermovoltage is given by the following expression [26]

$$V_{\text{th}} = \left[S(V_{\text{gate}} + \Delta V_{\text{zero}}) - S(V_{\text{gate}})\right] \cdot \Delta T = \Delta S \cdot \Delta T, \quad (2)$$

with a typical value of $\Delta V_{\text{zero}} \sim 9V$ (Figure 4d). We further determine a local maximum temperature increase of $\Delta T = 3$ K for the used laser power from finite element simulations. In turn, we can compute a gate voltage dependence of the thermovoltage $V_{\text{th}}$ according to equation (2) (line in Figure 4g). The computed dependence qualitatively captures both the experimentally observed ambipolar sign change of the photocurrent at $V_{\text{gate}} \sim 0V$ as well as a higher photocurrent amplitude for negative $V_{\text{gate}}$ (open circles in Figure 4g). Hereby, the model suggests that a photo-thermoelectric current within the surface states dominates the photocurrent in the constriction. Along this line, the values of $V_{\text{th}} \sim 1.2$ μV and $I_{\text{photo}} \sim 200$ pA in Figure 4g yield a resistance $R \sim 6$ kΩ, which is consistent with the local resistance at the laser spot. For $V_{\text{gate}} < 50$ V (Figure 4g), our model clearly starts to deviate from the experimental data. In this regime, the bulk valence band states increasingly contribute to the photocurrent, which are neglected in the calculation. For each laser position along the constriction, the sign of $\Delta V_{\text{zero}}$ can switch (Figure 4d). In turn, $V_{\text{th}}$ switches polarity according to equation (2). This explains the polarity change of the photocurrent for certain positions along the constriction (Figures 4b and 4c). We point out that the observed sign change excludes a photovoltaic effect to dominate the photoresponse of the bottom surface states. This can be seen by the following simple argument. Let us assume that at a negative gate voltage, the chemical potential fluctuations give rise to a doping profile of p-p$^+$-p along the constriction. When the global back-gate is tuned to positive voltages, it shifts the doping profile finally to a profile of n$^-$-n-n$^-$. For all gate voltages, however, the direction of the local built-in fields and accordingly the polarity of a local photovoltaic current would remain unchanged for a certain laser position [39]. With this argument, the photovoltaic effect can only influence the variation of $V_{\text{zero}}$ at each laser position along the constriction (Figure 4d), but it does not explain the ambipolar dependence vs $V_{\text{gate}}$ in Figure 4c.

In our interpretation, the laser spot defines an extrinsic length scale of the optoelectronic dynamics in the (Bi$_{0.5}$Sb$_{0.5}$)$_2$Te$_3$-films. However, the occurrence of micrometer sized photocurrent patterns suggests long-range potential fluctuations within the bottom surface state on a comparable length scale. As argued above, we interpret them to stem from the interaction with the SrTiO$_3$-substrate and defects at the edges of the (Bi$_{0.5}$Sb$_{0.5}$)$_2$Te$_3$-films. Moreover, the laser spot probes a certain number of current stream lines that connect to the macroscopically distant contacts [33, 34]. Then, the AFM-defined constrictions confine all current stream lines on a lateral width which compares well with the laser spot diameter. In this sense, the constrictions are quasi-one-dimensional, and there is always one dominating potential fluctuation and therefore, current direction per laser position. In turn, we can measure a zero-crossing of the photocurrent amplitude vs. gate voltage (Figure 4c). In the center of the (Bi$_{0.5}$Sb$_{0.5}$)$_2$Te$_3$-films, the excitation spot generates a photocurrent isotropically, i.e. also into directions which are perpendicular to the current stream lines. Hereby, further current stream lines possibly outside of the laser spot are involved, which then connect equally to source and drain. This explains that the corresponding photocurrent patterns rather do not change their polarity as a function of gate voltage (Figures 3a, 3b, and 3c). Consistently, this two-dimensional photocurrent response is maximum at the charge neutrality point (Figure 3d). As argued in the last section, a local photovoltaic effect is also consistent with the fact that the photocurrent patterns rather do not change their polarity vs. gate voltage. For the lateral edges of the $(Bi_{0.5}Sb_{0.5})_2Te_3$-films, the data suggest that the involved states are energetically lowered (Figure 4e). Finally, we note that the cooling cycles can slightly shift the voltage $V_{min}$ of the minimum conductance (as defined in Figure 3d), which can be considered approximately as the gate voltage for the charge neutrality point. However, the distinction between optoelectronically quasi-one-dimensional and quasi-two-dimensional responses is independent of the temperature cycling.

In summary, we use spatially resolved photocurrent measurements to probe the electronic states at the interface of $(Bi_{0.5}Sb_{0.5})_2Te_3$-films and the underlying $SrTiO_3$-substrate. The measured photocurrent patterns are consistent with the existence of submicron chemical potential fluctuations near the topological insulator/substrate interface. We introduce an AFM-based nano-plowing lithography to locally thin the $(Bi_{0.5}Sb_{0.5})_2Te_3$-films and to laterally define constrictions with single potential fluctuations. For the lateral constrictions, the photocurrent is consistent with a dominating photo-thermoelectric effect within the surface states of the topological insulator. We extract potential fluctuations on the order of meV. Our results may prove essential for the design and fabrication of nanoscale circuits from topological insulators.

## Supplementary data

Supplementary data available: Growth parameters. AFM lithography. Time-integrated photocurrent measurement.

## Author information

Corresponding Author * (A.W.H.) E-Mail: holleitner@wsi.tum.de

The authors declare no competing financial interest.

## Acknowledgements

We gratefully acknowledge financial support of the DFG-grant HO 3324/8 of the SPP 1666 on topological insulators, the Center for NanoScience (CeNS) and the Munich Quantum Center (MQC).

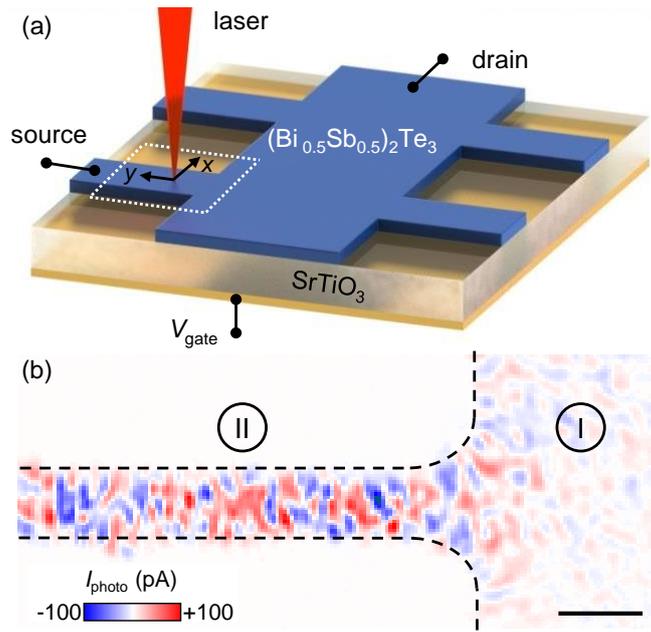

**Figure 1.** Local photocurrents in topological insulators. (a) Microstructured thin film of $(Bi_{0.5}Sb_{0.5})_2Te_3$ on $SrTiO_3$-substrate. Photocurrent is measured between unbiased source and drain contacts. The charge carrier density is adjusted via the global backgate voltage $V_{gate}$. (b) High resolution photocurrent map of the region marked by the rectangle in (a). Red (blue) denotes a current from source (drain) to drain (source). Dashed lines indicate sample boundary. Scale bar is 10 μm. Experimental parameters are $P_{laser}$ = 70 μW and $T_{bath}$ = 77 K.

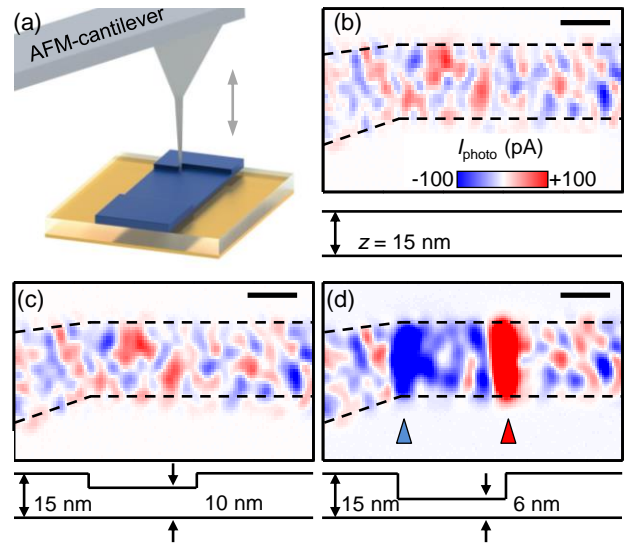

**Figure 2.** Dynamic plowing lithography. (a) The tip of an atomic force microscope (AFM) is used for mechanical milling of a $(Bi_{0.5}Sb_{0.5})_2Te_3$-film. (b) Photocurrent map of pristine material with thickness 15 nm. Dashed lines indicate sample boundary. (c) Photocurrent map after the film is thinned by 5 nm. Characteristic photocurrent patterns remain unchanged. (d) Photocurrent map after the film is thinned by 9 nm in total. At vertical step edges, the photoresponse is enhanced (triangles). Colorscale is clipped for contrast. Scale bars are 5 µm. Lower panels depict schematic side views of the film with an indicated thickness after each AFM-step. Experimental parameters are $V_{gate}$ = 0 V, $P_{laser}$ = 65 µW, and $T_{bath}$ = 77 K.

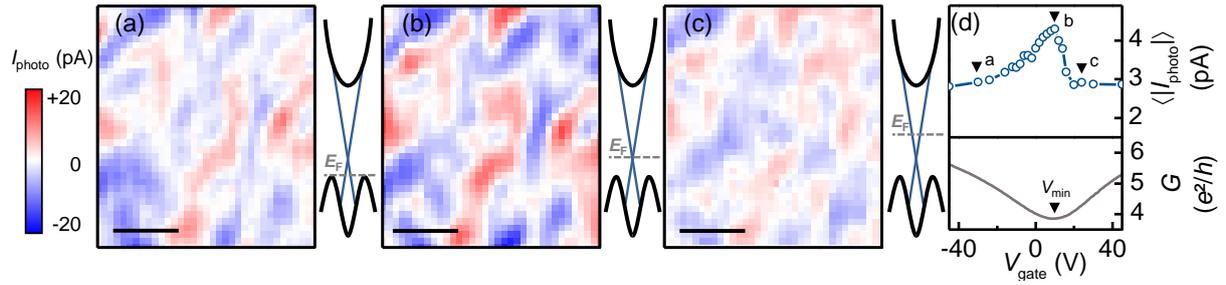

**Figure 3.** Photocurrents at the charge neutrality point. (a) Photocurrent map for $V_{gate}$ = -30 V, (b) +10 V, and (c) +24 V. Scale bars are 2 µm. Insets sketch the position of the Fermi-level in the bottom surface state. (d) Absolute value of photocurrent $\langle |I_{photo}| \rangle$ averaged over two-dimensional maps vs. gate voltage (top), and corresponding sheet conductance $G$ (bottom). Letters a, b, and c indicate measurements of (a), (b), and (c). $V_{min}$ highlights the minimum of the conductance $G$. Experimental parameters are $P_{laser}$ = 70 µW and $T_{bath}$ = 4.2 K.

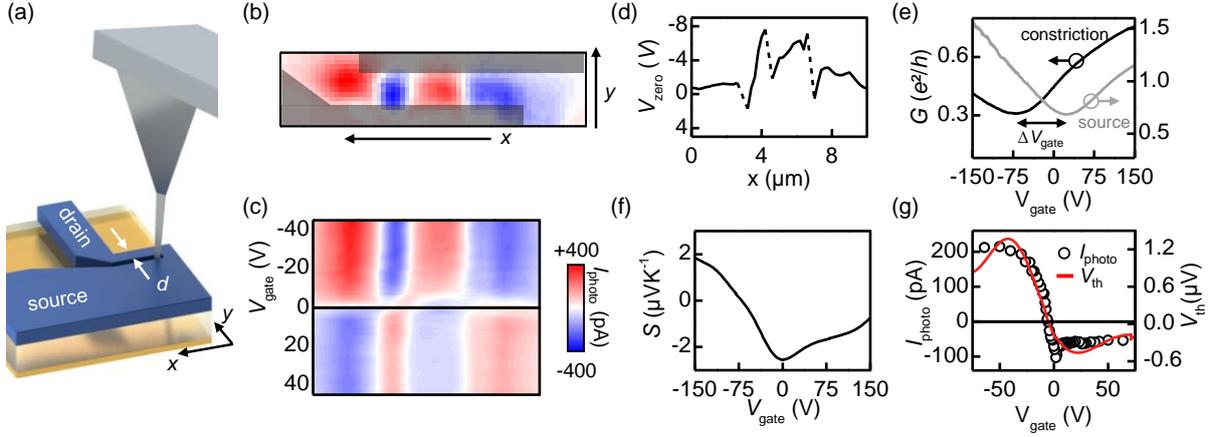

**Figure 4.** Long-range electrostatic potential fluctuations and thermoelectric photocurrent generation. (a) AFM lithography of a 1.8 µm wide constriction. (b) Photocurrent map of the nano-plowed constriction at $V_{gate}$ = -40 V with sequence of single potential fluctuations. Shaded areas indicate lithography and boundary. (c) Photocurrent vs. $V_{gate}$ along the constriction. The photocurrent changes sign for all positions around a varying $V_{zero}$ close to -2 V. (d) Variations of $V_{zero}$ along the length of the constriction extracted from (c). (e) The dark conductance through the constriction (black line) is shifted in $V_{gate}$ compared to the dark conductance of the source contact (gray line). (f) Seebeck coefficient $S$ calculated from equation (1) for the constriction. (g) Thermovoltage $V_{th}$ calculated from equation (2) and measured photocurrent as function of gate voltage. The thermovoltage and the photocurrent change polarity at $V_{gate}$ ~ -2 V. Experimental parameters are $P_{laser}$ = 70 µW and $T_{bath}$ = 4.2 K.

# Chemical Potential Fluctuations in Topological Insulator (Bi$_{0.5}$Sb$_{0.5}$)$_2$Te$_3$ Films Visualized by Photocurrent Spectroscopy

Christoph Kastl, Paul Seifert, Xiaoyue He, Kehui Wu, Yongqing Li, Alexander Holleitner

- Supplementary data -

**Supplementary Note 1: Growth and lithography parameters of (Bi$_{0.5}$Sb$_{0.5}$)$_2$Te$_3$ films.**

The (Bi$_{1-x}$Sb$_x$)$_2$Te$_3$ thin films are grown in a home-made molecular beam epitaxy (MBE) system with a base pressure better than $1 \times 10^{-10}$ mbar. Prior to the growth, the SrTiO$_3$(111) substrates are boiled in deionized water at 85 °C for 50 minutes, then annealed in pure O$_2$ environment at 1050 °C for 2 hours in order to obtain smooth surfaces. During the growth, high purity Bi (99.999%), Sb (99.999%) and Te (99.999%) are evaporated from Knudsen cells. A quartz crystal thickness monitor is used to calibrate the flux rate. A Bi/Te flux ratio (about 1:10) was kept to ensure Te-rich growth condition. The Bi:Sb compostion ratio was varied by adjusting Sb flux. The deposition rate is about 0.25 nm/min. The substrate temperature is maintained at 360 °C throughout film growth. More details of the the sample growth can be found in Ref. [1]

After being taken out the MBE chamber, the (Bi$_{1-x}$Sb$_x$)$_2$Te$_3$ films are patterned into 50 μm wide Hall bars by standard photolithography, followed by reactive ion etching with an Ar flow rate of 40 sccm and a coil power of 80 W. The etching rate is about 1 nm/s. Cr/Au (3 nm/30 nm) layers are then deposited on the back of the SrTiO$_3$ substrates and the top surfaces of the (Bi$_{1-x}$Sb$_x$)$_2$Te$_3$ thin films to serve as back-gate electrodes and ohmic contacts, respectively.

**Supplementary Note 2: Atomic force microscope lithography on (Bi$_{0.5}$Sb$_{0.5}$)$_2$Te$_3$ films.**

We use the tip of an atomic force microscope (AFM) and employ a so-called dynamic plowing lithography to thin down the (Bi$_{0.5}$Sb$_{0.5}$)$_2$Te$_3$-films. In tapping mode, the external drive amplitude of the AFM-tip is increased and the set-point of the feedback loop is decreased such that the interaction between the tip and the films is sufficient to mechanically remove sample material from the top surface of the (Bi$_{0.5}$Sb$_{0.5}$)$_2$Te$_3$. Depending on the set-point different modes of operation are possible, as shown in Supplementary Figure 1 and Supplementary Table 1. Imaging and lithography are done without exchanging the tip. For the presented data, we use an Asylum MFP-3D atomic force microscope (AFM) in combination with a diamond coated tip (DCP11 by NT-MDT). The device parameters for different task are listed in Supplementary Table 1. Note that the parameters are specific to the used MFP-3D AFM.

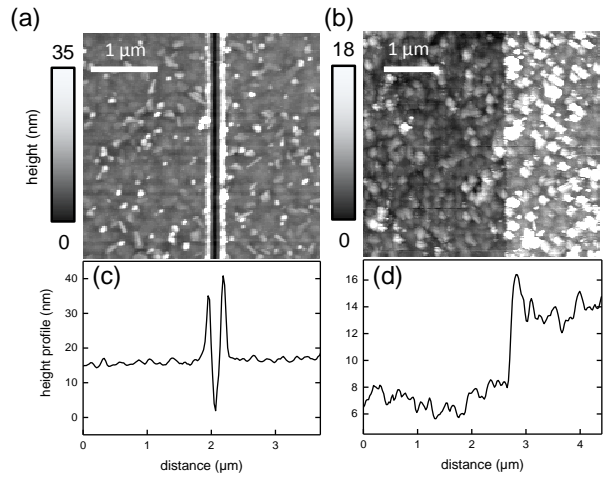

**Supplementary Figure 1.** Dynamic plowing lithography on $(Bi_{0.5}Sb_{0.5})_2Te_3$ films. (a) Atomic force microscope (AFM) image of a nano-plowed, 15 nm deep trench. The width of the trench is 140 nm. (b) AFM image of a nano-plowed step edge. The step height is 5 nm. (c) and (d) are horizontal line cuts.

| Parameter | Imaging | Thinning (step edge) | Cutting (trench) |
|---|---|---|---|
| **Drive amplitude** | 10 mV | 1000 mV | 1000 mV |
| **Set-point** | 730 mV | 100 mV | 5 mV |

**Supplementary Table 1** Device parameters for dynamic plowing lithography.

## Supplementary Section 3: Optoelectronics of AFM-fabricated constrictions

Supplementary Figure 2 depicts photocurrent measurements at a constriction, which is fabricated at the center of a Hall-bar by AFM-lithography only (Supplementary Figure 2a and 2b). The constriction is fabricated without reactive plasma etching. Interestingly, we observe a characteristic polarity change of the photoresponse at the constriction when sweeping the gate voltage (Supplementary Figure 2c). Hereby, the results on this AFM-defined constriction are similar to the data presented in the context of Figures 4 of the main manuscript. The sign change is characteristic for a band bending near lateral boundaries due to possible defect states by the nanofabrication process. In the constriction presented in the Supplementary Figure 2, all edges of the constrictions are fabricated by the AFM lithography, whereas the constriction presented in the main manuscript is fabricated on an edge of the $(Bi_{0.5}Sb_{0.5})_2Te_3$-film which is created by reactive plasma etching beforehand. Hereby, we conclude that the occurrence of defect states and their influence on the optoelectronic properties of a topological insulator are generic to nanofabricated circuits.

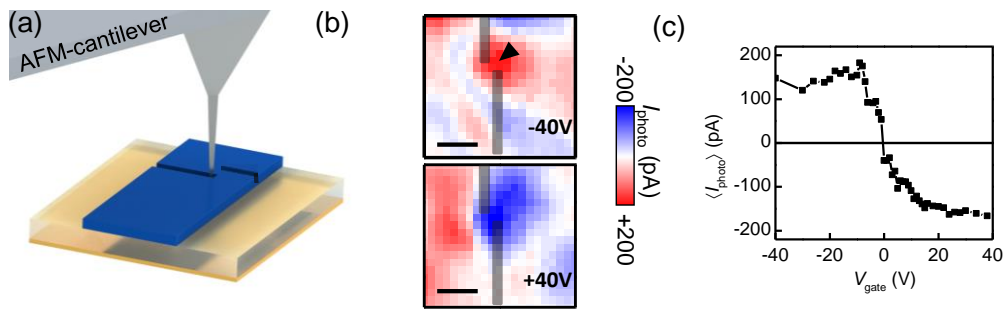

**Supplementary Figure 2.** Dynamic plowing lithography on $(Bi_{0.5}Sb_{0.5})_2Te_3$ films. (a) A constriction is formed by plowing two lines with a width of 250 nm at a gap distance of 200 nm. (b) Photocurrent map of the constriction for a gate voltage of $V_{gate}$ = -40 V (top panel) and $V_{gate}$ = +40 V (bottom panel). The gray areas sketch the position of the nano-plowed trenches forming the constriction. Scale bars are 1 μm. The triangle indicates the photocurrent signal at the constriction. (c) Photocurrent at the constriction vs. gate voltage changes polarity around $V_{gate}$ ~ 0 V. Experimental parameters are $P_{laser}$ = 62 μW and $T_{bath}$ = 4.2 K.